# How should fishing mortality be distributed under balanced harvesting?

Michael J. Plank[1]

1. School of Mathematics and Statistics and Te Pūnaha Matatini, University of Canterbury, Christchurch 8140, New Zealand.

Email: michael.plank@canterbury.ac.nz

**Short title:** Fishing intensity under balanced harvesting

**Abstract**
Zhou and Smith (2017) investigate different multi-species harvesting scenarios using a simple Holling-Tanner model. Among these scenarios are two methods for implementing balanced harvesting, where fishing is distributed across trophic levels in accordance with their productivity. This note examines the effects of a different quantitative implementation of balanced harvesting, where the fishing mortality rate is proportional to the total production rate of each trophic level. The results show that setting fishing mortality rate to be proportional to total production rate, rather than to productivity per unit biomass, better preserves trophic structure and provides a crucial safeguard for rare and threatened ecological groups. This is a key ingredient of balanced harvesting if it is to meet its objective of preserving biodiversity.

**Keywords:** Ecosystem approach to fisheries; ecosystem-based fisheries management; Holling-Tanner model; multi-species harvest strategies; productivity; trophic structure.

## 1. Introduction

Balanced harvesting (BH) is a proposed approach to fishing, which "distributes a moderate mortality from fishing across the widest possible range of species, stocks, and sizes in an ecosystem, in proportion to their natural productivity" (Garcia et al., 2012). BH was developed as a strategy to conserve biodiversity and ecosystem function, consistent with the concept of ecosystem-based fisheries management (Zhou et al., 2010). Among the hypothesised benefits of BH are improved biodiversity conservation, reduced disruption of community structure, increased resilience to fishing and increased biomass yields (Law et al., 2012; Charles et al., 2015; Garcia et al., 2015). BH has been criticised as being difficult to implement, implying the harvest of species of conservation concern such as seabirds and marine mammals, and reducing economic profits from fishing by shifting catches towards species and/or sizes with low market value (Burgess et al., 2015; Froese et al., 2015; Pauly et al., 2016; Reid et al., 2016). Scientific studies can help us to understand the consequences of alternative fishing policies and harvest rules. Assessing the relative value of those outcomes involves a complex set of trade-offs among ecological, economic and societal values, and is ultimately a sociopolitical judgement rather than purely a scientific one. However, this judgement needs to be informed by the best possible scientific evidence and mathematical models have an important role to play in this.

BH has been studied using size-based community models (Law et al,. 2012, 2015; Jacobsen et al., 2014; Kolding et al., 2015) and using multi-species ecosystem models, such as Ecopath and Atlantis (Bundy et al., 2005; Garcia et al., 2012; Kolding et al., 2016; Heath et al., 2017). These studies have shown that BH has the potential to maintain or increase total sustainable ecosystem yield, albeit consisting of a greater proportion of species and sizes of low commercial value, while better preserving ecosystem structure. However, these models are relatively complex and their results may be sensitive to model assumptions or noisy data. On the other hand, simple models can sometimes offer qualitative insights that more complex models cannot. Given the controversy generated by BH, it is appropriate that its consequences be investigated using a range of different modelling approaches (Garcia et al., 2014).

Zhou and Smith (2017) investigate the effects of various types of selective or non-selective fishing in a simple, equilibrium Holling-Tanner model (Tanner, 1975) of a fish community split into three trophic levels (TLs). In this model, a fishing scenario must specify not only the overall intensity of fishing, but also how fishing mortality is distributed across TLs. Among the fishing scenarios considered by Zhou and Smith (2017) are two forms of BH, in which fishing mortality rate $F$ is proportional either to the current productivity or to the maximal productivity of each TL. Productivity is defined as the amount of new biomass produced per unit of existing biomass per unit time, with dimensions time$^{-1}$ (Garcia et al., 2012) and denoted $P/B$ in Ecopath models (Christensen and Pauly, 1992). Maximal productivity is the productivity at close-to-zero biomass, which, under the assumptions of the Holling-Tanner model, is equivalent to the intrinsic rate of increase, $r$. These two fishing scenarios are referred to as $F{\sim}P/B$ and $F{\sim}r$, respectively.

Zhou and Smith (2017) calculate the biomass, yield and disruption of trophic structure resulting from each fishing scenario examined and show that the only scenario that perfectly preserves trophic structure is fishing exclusively on the lowest TL (representing planktivorous fish). The reason is that, under the assumptions of the Holling-Tanner model, the biomass depletion of the lowest TL is transmitted up the food chain, causing proportional biomass depletions of the unfished higher TLs. The obvious downside to this fishing scenario is that the catch is exclusively from the lowest TL, which may not be economically desirable. In contrast, both forms of BH examined by Zhou and Smith (2017) ($F{\sim}P/B$ and $F{\sim}r$) provide a catch composed of all three TLs and a higher total yield. However, both scenarios also cause significant disruption to the trophic structure, with disproportionate depletion of the higher TLs. In addition, under fishing with $F{\sim}P/B$, there is a sudden collapse of all three TLs as the exploitation ratio (ratio of yield to production rate, which is the control parameter in setting $F{\sim}P/B$) is increased from 0.85 to 0.95.

An alternative strategy for BH, not considered by Zhou and Smith (2017), is to set the fishing mortality rate $F$ to be proportional to the total production rate $P$ of each TL (dimensions mass × time$^{-1}$).

This strategy has been investigated previously in size-spectrum models (Law et al., 2015). A key feature of this approach is that it incorporates a density dependence into the fishing mortality rate. This means that, as an ecological group, such as a species or TL, becomes depleted and its total production rate drops, the fishing mortality rate on that group is automatically reduced. Since, in an equilibrium model, yield $Y$ is equal to fishing mortality rate $F$ multiplied by biomass $B$, fishing with $F \sim P/B$ is equivalent to setting a constant exploitation ratio ($Y/P$) across all TLs (Kolding et al., 2016). Fishing with $F \sim P$ means that $Y \sim PB$, so this calls for a higher exploitation ratio on TLs with higher biomass (Heath et al., 2017). This note investigates the effect of BH with $F \sim P$ in the model considered by Zhou and Smith (2017).

## 2. Methods

The Holling-Tanner model considered by Zhou and Smith (2017) is defined by the following differential equations for the biomass $B_i$ of TL $i$ ($i = 1, 2, 3$):

$$\frac{dB_i}{dt} = r_i B_i \left(1 - \frac{B_i}{K_i}\right) - M_i B_i - F_i B_i, \qquad (1)$$

where $r_i$ is the intrinsic growth rate, $K_i$ is the carrying capacity, $M_i$ is the natural mortality rate and $F_i$ is the fishing mortality rate of TL $i$. The carrying capacity of TL1 is constant. The carrying capacities of TL2 and TL3 are given by $K_i = e_{i-1,i} B_{i-1}$, where $e_{i-1,i}$ is the efficiency of biomass transfer from TL $i-1$ to TL $i$. The natural mortality rates for TL1 and TL2 are given by a type-II function of predation by the TL above: $M_i = \frac{p_{i,i+1} B_i B_{i+1}}{D_i + B_i}$. The natural mortality rate for TL3 is constant. All parameter values are the same as those used by Zhou and Smith (2017).

To set the fishing mortality rate in proportion to the production rate ($F \sim P$), define

$$F_i = c r_i B_i \left(1 - \frac{B_i}{K_i}\right), \qquad (2)$$

where $c$ is a constant controlling the overall level of fishing intensity. Under the scenarios considered by Zhou and Smith (2017), overall fishing intensity is controlled by a constant of proportionality $f$ that is dimensionless and can be varied between 0 and 1. The constant $c$ in Eq. (2) has dimensions mass$^{-1}$ and does not have an a priori defined range. Values of $c$ are trialed in the model to find an appropriate range encompassing the maximum sustainable yield for each TL.

The differential equations, Eq. (1), for each TL are solved numerically until an equilibrium is reached. The equilibrium (i.e. sustainable) yield for TL $i$ is calculated as $Y_i = F_i B_i$. Following Zhou and Smith (2017), disruption of trophic structure is measured in two ways: (i) the slope of the relationship between TL and log biomass; (ii) the disturbance index (DI), which is based on biomass ratios of adjacent TLs (Bundy et al., 2005). This process is repeated for a range of values of the constant $c$ that determines the overall intensity of fishing.

## 3. Results

Fig. 1 shows the effect of fishing in proportion to production rate ($F \sim P$) on the relative biomass and yield of each TL (compare with Figs. 1-6 of Zhou and Smith, 2017), for values of the overall fishing intensity $c$ ranging from 0 to 0.01 per unit biomass. Fig. 2 shows the total ecosystem biomass and yield against average fishing mortality rate $F$ (compare with Fig. 7 of Zhou and Smith, 2017). For comparison, Fig. 2 also shows the two BH scenarios examined by Zhou and Smith (2017) and the scenario where fishing is only on TL1. From these results, several observations are possible for this simple model:

1. Fishing with $F \sim P$ preserves trophic structure almost perfectly (the three TL biomass curves are almost indistinguishable in Fig. 1a, and the slope and DI are barely affected by fishing in Fig. 1c). This is mainly because this strategy focuses most of the fishing on the lowest TL, where the production rate is highest. As with the strategy of fishing only the lowest TL, the biomasses of the higher TLs are depleted in proportion as a result of reduced carrying capacities. These TLs have relatively low biomasses and hence production rates so, under $F \sim P$, they are subjected to relatively low fishing mortality rate. This protects them from the disproportionate biomass depletions that they suffer under $F{\sim}P/B$.
2. The differences in the maximum ecosystem sustainable yield (MESY, maxima of curves in Fig. 2b) among the different fishing scenarios are small. The MESY from fishing with $F{\sim}P$ is 0.4% higher than fishing on TL1 only, 6% lower than $F{\sim}P/B$ and 8% lower than $F{\sim}r$.
3. At least 95% of the yield from fishing with $F{\sim}P$ comes from the lowest TL. However, this is largely a consequence of the simple model assumption that the carrying capacity of TL $i$ is 20% of the biomass of TL $i-1$. In reality, biomass is distributed more equally across TLs (Sheldon et al., 1972; Boudreau and Dickie, 1992; Andersen and Beyer, 2006) and in that case fishing with $F{\sim}P$ will produce a significant yield from multiple TLs. This is a clear advantage over the strategy of fishing exclusively on the lowest TL, which will obviously never produce any catch from higher TLs.
4. Because of the inbuilt density dependence of fishing with $F{\sim}P$, there is no sudden collapse of the biomass as the control parameter $c$ is increased, but instead a gradual biomass decline once MESY is exceeded (compare Figure 1 with Figure 6 of Zhou and Smith, 2017).

## 4. Discussion

Ecosystem-based fisheries management involves a range of trade-offs among competing objectives, including maximising yields, maximising economic profits, and minimising environmental impact and biodiversity loss (Zhou et al., 2010; Burgess et al., 2015; Charles et al., 2015; Jacobsen et al., 2016). How these objectives should be weighted is more of a sociopolitical question than a scientific one, and will clearly involve compromises. For example, strictly maximum ecosystem yield is likely to come from the extreme strategy of harvesting a single, low TL species and deliberately driving all competing or predatory species to extinction. At the opposite extreme, minimising ecosystem disruption would involve cessation of all fishing activity, which would clearly have dire consequences for world food supply.

BH is a possible strategy for reconciling some of these objectives, but quantitative implementations of BH in modelling studies differ (Law et al., 2012; Jacobsen et al., 2014; Kolding et al., 2016; Plank, 2016; Heath et al., 2017). Zhou and Smith (2017) used a Holling-Tanner model to investigate the effect of two potential implementations of BH in which fishing mortality rate $F$ is proportional to either the maximal productivity $r$ or the current productivity (i.e. production per unit biomass, $P/B$) of each TL. The current study compares these strategies with a third BH strategy in which fishing mortality rate is proportional to the total production of each TL ($F{\sim}P$). Under the Holling-Tanner model, fishing with $F{\sim}P$ preserves trophic structure better than either of the BH strategies considered by Zhou and Smith (2017) and has a similar outcome to fishing only on the lowest TL.

The key result from the current study is not the outcome of fishing with $F{\sim}P$ under the specific Holling-Tanner model, but rather that a strategy that reduces fishing mortality rate at low biomass is an essential ingredient if BH is to meet its objective of preserving ecosystem structure. Setting fishing mortality rate in proportion to the productivity $P/B$ of an ecological group potentially leaves less abundant groups vulnerable to overfishing. In contrast, setting fishing mortality rate in proportion to the total production $P$ of an ecological group reduces fishing pressure on less abundant groups and provides an inbuilt safeguard against fishing-induced collapse. There is an intuitive reason for this: as an ecological group is driven to low abundance, $P$ and $B$ tend to zero, but productivity $P/B$ can remain almost constant as $P$ and $B$ decline in proportion to each other.

Fishing with $F{\sim}P$ is one strategy that reduces fishing mortality on rare or threatened groups, but there may be others. Harvest control rules, for example, often prescribe a fishing mortality rate that reduces in proportion to the biomass of an individual stock (Kvamsdal et al., 2016). Fishing with $F{\sim}P$ provides a simple strategy for extending this simple concept to a BH framework across multiple ecological components.

BH calls for a moderate fishing mortality to be distributed across ecological groups (Garcia et al.,

2012). However, the meaning of "moderate" is left undefined, may be context- and ecosystem-dependent (Heath et al., 2017) and will, in general, involve a trade-off between increasing ecosystem yield and reducing environmental impact. Some ecological groups will be intrinsically more sensitive to the effects of fishing than others, so a given fishing mortality might be moderate for one group but extreme for another group that has a similar *P/B* but lower *B*. Fishing with *F~P/B* will treat these two groups the same, whereas fishing with *F~P* helps to safeguard rare or vulnerable groups by reducing fishing pressure at low biomass.

The equilibrium Holling-Tanner model is clearly very simplistic compared to a real ecosystem. As Zhou and Smith (2017) point out, testing of alternative fishing strategies and harvest rules needs to be carried out under a range of different model frameworks and structural assumptions to ensure that robust conclusions are reached. The results in this study show that that setting *F~P* keeps trophic structure much closer to its natural state than does *F~P/B* in the Holling-Tanner model. This is consistent with results comparing these two harvest rules in a more complex size-spectrum model (Law et al., 2014), but it would be beneficial to further investigate the sensitivity of this finding to specific model assumptions.

BH calls for the harvesting of previously unexploited taxa or groups currently regarded as bycatch. There are clearly risks associated with this strategy and it is essential that it is implemented in a way that prioritises biodiversity. Setting fishing mortality rate in proportion to the current production rate provides a more precautionary approach to conserving biodiversity than setting fishing mortality rate in proportion to the productivity per unit biomass, because of its inbuilt safeguard for rare or threatened ecological groups. This is consistent with the aim of BH of "maintaining the relative size and species composition" (Garcia et al., 2012) and with the requirements of the Convention on Biological Diversity (UNEP/CBD, 1998) for conservation of ecosystem structure and functioning (Garcia et al., 2015).

**Acknowledgements**
The author is grateful to Shijie Zhou, Richard Law, Serge Garcia and Jeppe Kolding for discussions about the effects of fishing in multi-species models.

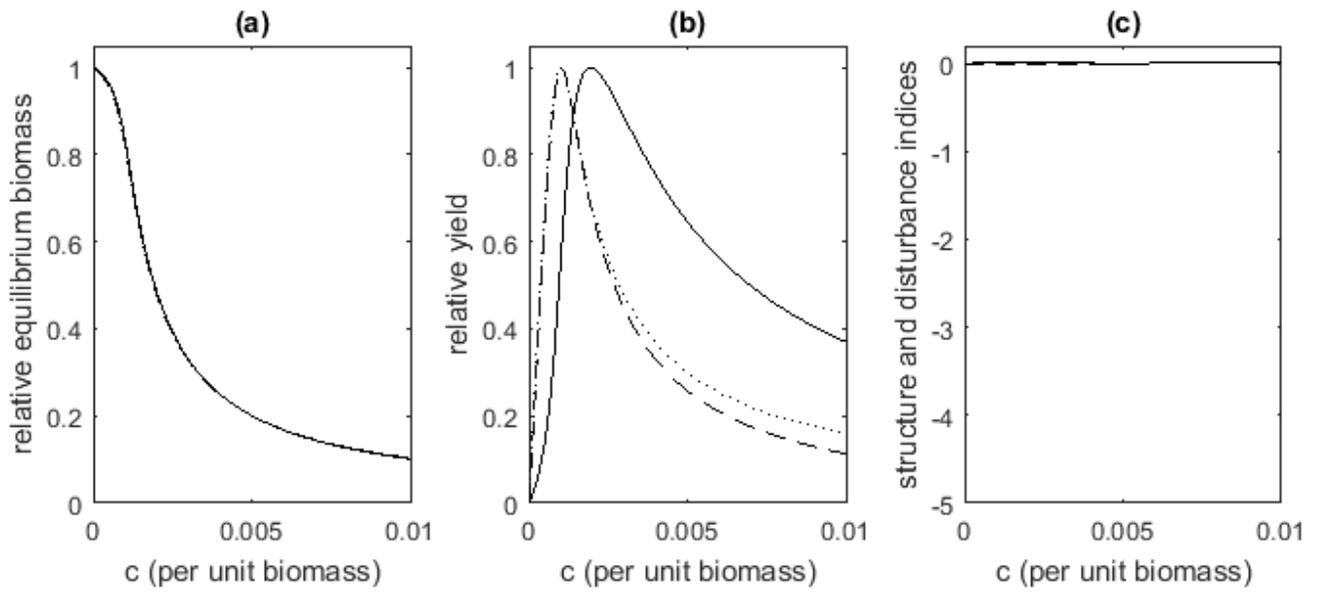

**Figure 1.** Equilibrium biomass, yield and trophic structure for fishing proportional to production rate ($F_i = cP_i$): (a) equilibrium biomass relative to maximum; (b) yield relative to maximum (TL1 solid, TL2 dashed, TL3 dotted); (c) slope of log biomass-TL relationship (dashed) and disturbance index (solid).

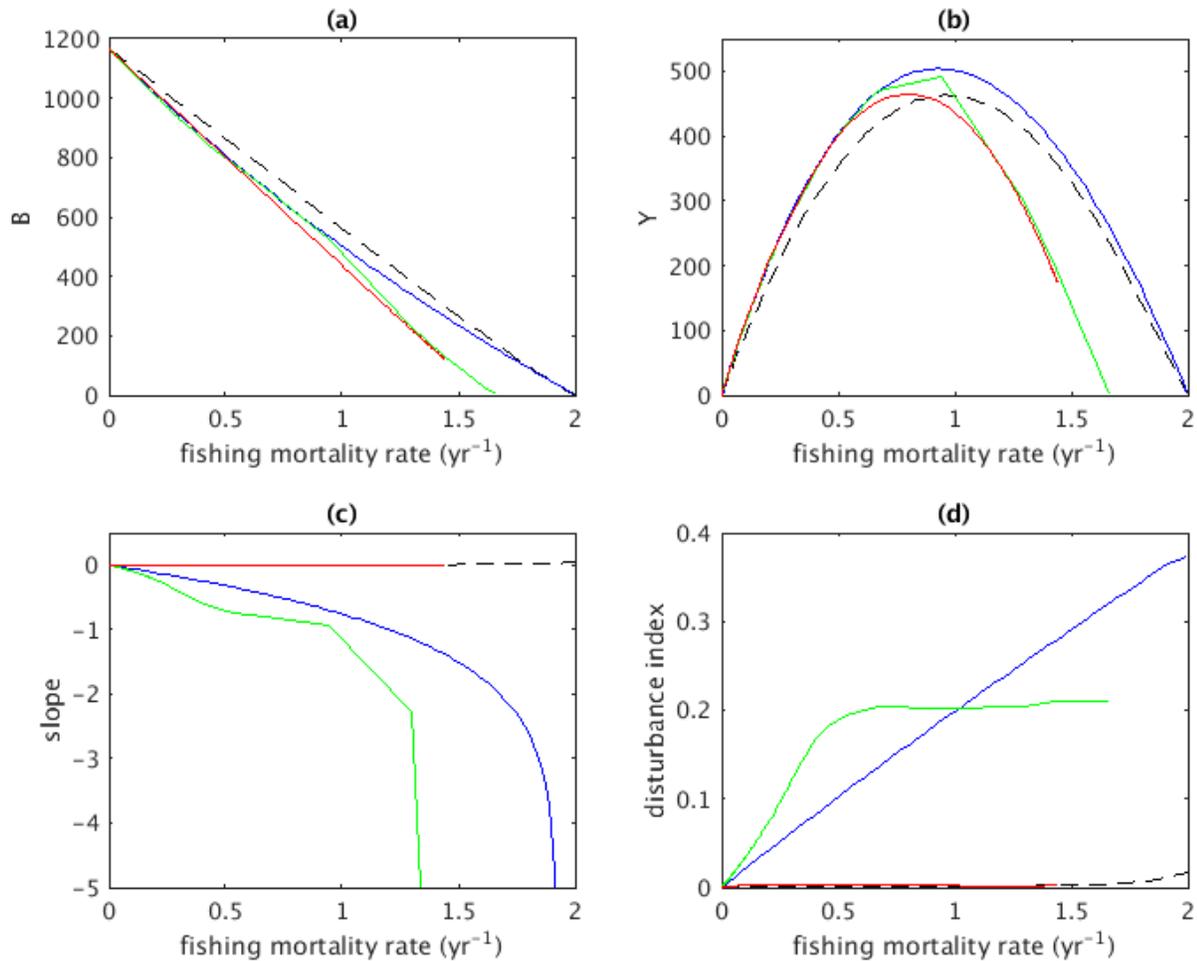

**Figure 2.** Comparison of (a) total biomass, (b) total yield, and trophic structure measured by (c) slope and (d) disturbance index, against fishing mortality rate across all trophic levels at equilibrium for alternative fishing scenarios: fishing TL1 only (black dashed); fishing with $F\sim r$ (blue solid); fishing with $F\sim P/B$ (green solid); fishing with $F\sim P$ (red solid). Fishing mortality rate is calculated as in Zhou and Smith (2017) as the total ecosystem yield divided by the total ecosystem biomass for scenarios that harvest all three TLs, and as the single TL yield divided by the single TL biomass for the scenario that only harvests one TL.